# Leader Cultural Intelligence and Organizational Performance

Saeed Nosratabadi [1], Parvaneh Bahrami [2], Khodayar Palouzian[3] Amir Mosavi [4,5,6*]

[1] Doctoral School of Management and Business Administration, Szent Istvan University, Godollo 2100, Hungary; saeed.nosratabadi@phd.uni-szie.hu
[2] Faculty of Management and Accounting, Department of Management, Allameh Tabatabai University, Tehran, Iran; bahrami_parvane@yahoo.com
[3] Faculty of Management and Accounting, Department of Management, Tehran University, Tehran, Iran; Email: palouzian@gmail.com
[4] Institute of Research and Development, Duy Tan University, Da Nang 550000, Viet Nam.
[5] Institute of Structural Mechanics (ISM), Bauhaus-Universität Weimar, 99423 Weimar, Germany
[6] Faculty of Civil Engineering, Technische Universität Dresden, 01069 Dresden, Germany
amir.mosavi@uni-weimar.de
*Corresponding Author: amir.mosavi@weimar-uni.de, amir.mosavi@kvk.uni-obuda.hu

**Abstract**

One of the challenges for international companies is to manage multicultural environments effectively. Cultural intelligence (CQ) is a soft skill required of the leaders of organizations working in cross-cultural contexts to be able to communicate effectively in such environments. On the other hand, organizational structure plays an active role in developing and promoting such skills in an organization. Therefore, this study aimed to investigate the effect of leader CQ on organizational performance mediated by organizational structure. To achieve the objective of this research, first, conceptual models and hypotheses of this research were formed based on the literature. Then, a quantitative empirical research design using a questionnaire, as a tool for data collection, and structural equation modeling, as a tool for data analysis, was employed among executives of knowledge-based companies in the Science and Technology Park, Bushehr, Iran. The results disclosed that leader CQ directly and indirectly (i.e., through the organizational structure) has a positive and significant effect on organizational performance. In other words, in organizations that operate in a multicultural environment, the higher the level of leader CQ, the higher the performance of that organization. Accordingly, such companies are encouraged to invest in improving the cultural intelligence of their leaders to improve their performance in cross-cultural environments, and to design appropriate organizational structures for the development of their intellectual capital.

**About the authors**

Saeed Nosratabadi is a Ph.D. candidate in management and business study and has published many research articles on different topics related to management and businesses such as human resource management, business model innovation, and marketing. Parvaneh Bahrami is a Ph.D. candidate in human resource management at the Department of Management, Allameh Tabatabai University, Tehran, Iran. Khodayar Palouzian is a Ph.D. candidate in Public Administration at the Department of Management, University of University, Tehran, Iran. Amir Mosavi is a professor of business intelligence at Obuda University, Hungary. He is the recipient of the Green-Talent Award, UNESCO Young Scientist Award, ERCIM Alain Bensoussan Fellowship Award, Endeavour-Australia Leadership Award, Humboldt Prize,

**Public Interest Statement**

One of the consequences of globalization is that the interaction among people with different cultural backgrounds has increased, and there are many organizations working in multinational environments. Cultural intelligence (CQ) is a skill required of the leaders of organizations working in cross-cultural environments to be able to communicate effectively in such environments. An organizational leader with a high level of CQ is able to have a better understanding of the quality of others' behaviors and mindsets in a multicultural environment. Therefore, this study is conducted to understand how leader CQ affects organizational performance. The results revealed that leader CQ directly and indirectly (i.e., through the organizational structure) affects organizational performance. In other words, in organizations that operate in a multicultural environment, the higher the level of leader CQ, the higher the performance of that organization.

1. **Introduction**

In today's post-industrial world, to achieve an immersive development that different societies are looking for, attention to soft skills sounds essential. In this regard, much research has been conducted on the cultural intelligence or quotient cultural (CQ) of leaders and its influencing factors. The organizations are confronted with emerging innovations and changes in the economic, social, technological, cultural, and political environments where the effective response in such a turbulent environment depends on the knowledge capabilities of the organizations. In addition, the advent of globalization and interdependence between countries increases the competition between organizations as well as the opportunities for business growth and development. Although the emerging phenomenon of globalization creates many business opportunities, these opportunities pose significant challenges that cultural conflict [1] is one of the most important ones. In fact, CQ deals with the understanding of cultural values and beliefs of various societies interacting with heterogeneous and diverse environments [2]. The competencies of leaders can play an important role in the success of organizations [3]. Using empirical research, Alzghoul et al. [4] show that leadership in an organization can have a positive impact on an organization's performance. Competent leaders with managerial abilities in multicultural environments can take the helm of leading organizations in such environments and facilitate the achievement of organizational goals.

Cultural diversity is one of the major issues the present organizations are facing with [5]. Unfortunately, most organizations tend to ignore cultural differences as an effective source of competitive advantage and avoiding thinking about cultural differences and the required skills to manage it. Under such circumstances, most experts believe that having such skillful leaders empowers the organizations to excel in the competition in global markets [6]. The organizations are required tools to improve the quality and use of these assets. One of these tools is leader cultural intelligence. The organizations need leaders who possess a set of intuitive and functional skills to successfully lead the organization in a dynamic and global environment [7, 8] because such leaders are able to anticipate rapid economic and cultural changes simultaneously with the rapid growth of global trade [9]. The research illustrates that the competitive advantage of an organization is related to the acquisition, the maintenance, and the use of strategic assets (both tangible and intangible assets), which in turn lead to a strong financial performance. Besides, organizations are successful in the knowledge-based economy that invest in opportunities resulted from intangible assets.

Although there is ample evidence in the literature that prove the predictive role of leader CQ in team performance [10, 11] and task performance [12, 13] in culturally diverse work teams, there is no study in the literature to empirically examine the contribution of leader CQ to organizational performance. In fact, there are few experimental studies that investigate the outcome and impact of the leader CQ on the organization. Therefore, the present study aims to address this gap in the literature and provides a better understanding of the crucial role of leader CQ in organizational performance. To develop such intelligent human resource, the organization should create an atmosphere in which the human resources are be able to perform their roles safely and comfortably that stimulate them to do their utmost for organizational goals. Organizational structure is one of the contextual and organizational-influencing factors which provides a foundation for growing intellectual capital [14, 15]. Understanding the structure of the organization is the beginning of any exploitation of existing interests and resources, presenting new combinations of existing resources, and ultimately paving the way for development and growth. Many dimensions and components have been presented in the literature for evaluating and designing organizational structure; among them, formalization, centralization, and complexity are the most referred dimensions of the organizational structure (e.g. [16-18]). Thus, the current study is conducted to investigate the impact of leader CQ on organizational performance mediated by organizational structure. In other words, this study offers empirical evidence to assess the elaboration ability of leader CQ to explain the organizational performance through organizational structure. First, the literature related to CQ, organizational performance, organizational structure, and the intersection of literature on organizational performance and cultural intelligence is reviewed and presented to develop the hypotheses of the study. Second, the methodology of the empirical study and data collection methods and data analysis procedures are described. Next, the results of testing the hypotheses are presented. Finally, the key results and theoretical and practical contributions of this study are discussed.

## 2. Research background

*2.1 Cultural intelligence*

The term of cultural intelligence was first coined in 2003 by two researchers, Earley and Ang, from the London School of Business [19]. They defined CQ as the ability of the individual to interact effectively with people who are culturally diverse with the cultural context of the individual [20]. CQ is the skill of managing people from diverse cultural backgrounds [1]. This intelligence consists of a set of skills that enable individuals to interact effectively with people from diverse cultures [21]. CQ is one of the dimensions of multiple intelligences and is similar in some respects to social intelligence and emotional intelligence, which focuses on a set of skills for effective behavior in different situations and its different with the other intelligence is that it refers to a set of cultural abilities [1]. Since CQ is a new concept, few studies have been done on this variable and its different dimensions. The division of CQ from the modern point of view was first put forward by researchers named Earley and Ang [19] where they divided CQ into four dimensions: cognitive, meta-cognitive, behavioral, and motivational CQ [19]. Following is part of previous research on these four dimensions:

The cognitive dimension is one of the main aspects of CQ that discusses having an empirical and cognitive context about patterns existing in a new cultural situation, which helps one process information better and more efficiently [22]. Influenced by this dimension, one strives to objectively and mentally acquire sufficient information about customs, traditions, and customs in diverse cultures and new patterns of behavior through learning or personal experience [23]. Thomas [2] argues that the cognitive dimension can be achieved by training one's own experience. The metacognitive dimension of CQ is related to the cognitive dimension in that one performs a mental process of the cognitive dimension obtained through personal experience or training, and then a particular understanding of cultural knowledge is created in one's mind [20] that includes strategic planning during a strategic interaction, monitoring the accuracy of its implementation during an interaction, and modifying mental patterns if deviated [1]. The behavioral dimension is another major aspect of CQ that refers to the appropriate reactions and behaviors during interactions with different cultures [22]. In this aspect of CQ, one is enabled to express appropriate and effective verbal and nonverbal behaviors during intercultural interaction, on the basis of the general judgment of the individual on new cultural environments [24]. The behavioral dimension does not restrict behavioral states in individuals but comprises the ability of individuals to adapt to customs, traditions, and lifestyles in different countries [25]. The motivational dimension of CQ refers to one's willingness to learn new cultural patterns and their behavior when entering an unfamiliar culture [21]. This dimension expresses one's ability to manage stress and psychological stress during interactions in new environments [22]. Besides, the role of external and internal stimuli in motivating one to adapt to culturally heterogeneous contexts is an important part of this dimension.

*2.2 Organizational Performance*

Nowadays, organizations seeking to achieve a source of competitive advantage by presenting high-quality service/products. Therefore, performance evaluation and quality improvement are essential. One of the tasks of managers is to monitor the performance of the organization. In general, though, it can be said that organizational performance is a broad concept that encompasses what the company produces and which areas it interacts with. In other words, organizational performance refers to how the organization reach its mission and performs its tasks and activities and the results of doing them [26].

Performance appraisal is one of the most effective tools in human resource management that, by applying this tool properly, not only the goals and missions of the organization will be achieved with the desired efficiency, but also the interests of the employees and the community can be reached[26]. The growth of the organization and the excellence of its staff will depend on having an effective evaluation system and applying its results. It is natural that developing and implementing a performance appraisal process can help an organization achieve its goals by enhancing employee effectiveness. There are two approaches to performance evaluation, one being the use of subjective criteria and the other is the use of objective criteria. Objective scales are a data-driven process that uses real figures of organizations, whereas subjective scales use respondents' perception [27]. Data Envelopment Analysis, Balanced Scorecard, Organizational Excellence Models, and European Foundation for Quality Management Model are subjective approaches proposed in the literature to performance evaluation. Data envelopment analysis is used as a mathematical programming method for evaluating decision-making units. This method has an initial assumption that decision-making units employ similar inputs to produce similar outputs. This approach is able to be used concurrently with respect to multiple and desirable inputs and outputs that have good production characteristics [28]. The Balanced Scorecard was introduced by a Harvard University Professor Robert Kaplan and David Norton, an Outstanding Advisors, in 1992, as an approach to align organizational performance measures with strategic goals and plans that improve management decision-making. The Balanced Scorecard provides managers with a formal framework for achieving a balance between financial and non-financial results in both the short and long term. It includes four perspectives, namely financial perspective, internal processes, customer, and growth and learning. This approach is a comprehensive measurement method for continuous improvement [29]. Self-assessment is a non-financial measure to evaluate the performance of organizations. Through this approach, managers evaluate operations, overall business insights, and continually improve their operations [30]. The emergence of excellence models began at the invitation of the Japan Institute of Scientists and Engineers (JUSE), Dr. Deming for a lecture on quality in Japan in 1950, and it was officially recognized in Japan in 1951 by the award of the Deming Prize. The European Quality Award was founded in 1991 with the EFQM model. In addition to the above models, there are many high-performance models developed by other countries, but the most popular ones are the Deming Prize, Malcolm Baldridge and European Quality Award models [31]. Other developed models have often been inspired by the three popular models mentioned above. Organizational excellence models by benchmarking successful companies around the world has been able to provide an appropriate framework for managing organizations in a competitive environment [32]. The distinctive feature of these models is the kind of attitude towards the organization (holism) that enables the manager to evaluate the organization under its authority and compare it with other similar organizations simultaneously. On the other hand, these models are usually designed in such a way that allows an organization to use different technologies. Following is the subjective model of the European Foundation for Quality Management.

In 1992, the European Foundation for Quality Management established the Quality Award on the basis of self-assessment [30]. The EFQM Excellence Model was an opportunity created by the fourteen leading European companies in 1998 with the creation of the European Foundation for Quality Management and the Evolution of Quality Systems with the support of the European Union [33]. This model has been increasingly used as a framework for evaluating performance and measuring the success of organizations in deploying new management systems. In addition, it is used as a common language to compare the performance and success of organizations.

The EFQM model constitutes nine main criteria, including five enabler criteria and four results measurements. According to this model, the performance enablers are: 1) leadership, 2) people, 3) strategy, 4) partnerships and resources, and 5) processes, products, and services. And four results measurements, which are: 1) people results, 2) customer results, 3) society results, and 4) business results [34]. Escrig-Tena, Garcia-Juan, and Segarra-Ciprés [34] argue that performance enablers affect the performance results, and the enablers determine the performance results quality. Using the EFQM model, as one of the most valid models of performance evaluation, despite some limitations, provides valuable opportunities in the organization for learning, balanced evaluation and evaluation of improvement opportunities. Using excellence models, organizations can measure their progress in implementing improvement programs at different times. This approach enables an organization to compare its performance with other organizations, especially with the best ones. In addition, the EFQM enables the organization to identify the differences between their existing and their desired situation and then based on this information to investigate the causes of their occurrence, determine the solution to optimize the situation and implement them [34].

H1. The performance enablers have a significant effect on the performance results

*2.3 Cultural intelligence and organizational performance*
With the development of the knowledge economy, intangible assets play an important role in the organization's strategies. CQ is a strategic, intangible, valuable, and irreplaceable resource that creates competitive advantage and better financial performance for the organization. This type of intelligence provides the knowledge and insight needed to promote social skills that enable the organizations to understand cultural differences and provide the human mental capacity to understand new information and create the ground for partnership [35]. Many executives around the world have come to realize that CQ is one of the most important components in gaining competitive advantage in today's knowledge-based economy, and effectively controlling this intelligence enables organizations to both dynamically and actively manage the intra-organizational aspects and to have successful inter-organizational relationships with the society and the other stakeholders. In addition, today, due to the rapid pace of environmental change, organizations need a strong corporate culture in terms of behavioral norms for intangible resources in comparison with the past, in order to be able to effectively and efficiently use their capitals in organizations.

CQ is a new domain of intelligence that has a great deal to do with diverse work environments. In fact, CQ focuses on the specific capabilities required for a high-quality personal relationship and effectiveness across different cultural contexts and allows employees to identify how others think and how to respond to behavioral patterns. As a result, it reduces intercultural communication barriers and gives individuals the power to manage cultural diversity. Diversity in culture leads to diversity in opinions. People who are culturally diverse have different perspectives, and the broader the range of ideas, the greater the chance of finding a good idea.

The behavioral dimension that is one of the key aspects of leader CQ includes the relationships the organization has with internal and external stakeholders, and it is a great challenge today where cultural diversity is ubiquitous and making appropriate communications are essential. More recently, if managers and staff in the organization wish to have effective communication in the organizational flow, the leaders of the organization must have an appropriate and varied CQ. Identifying, valuing, and supporting these differences can maximize employee productivity at work [36]. Hence, it can be interpreted that leader CQ has a positive and significant effect on the performance of employees and managers. On the other hand, for the workgroups to function effectively, the workforce itself must develop CQ. By increasing the CQ of staff, team members can build a foundation for mutual understanding and respect and enhance individuals' ability to identify solutions to their problems. As a result, if the organization's managers have high CQ, they will be better at selecting and using human resources than other organizations. As a result, CQ has a positive and significant impact on the human resources of the organization. Besides, the organizations need a good mix of all aspects of CQ to achieve higher performance. Leader CQ can create the best value for the organization by combining, deploying, integrating, and interacting with the dimensions as well as managing the flow of knowledge between them. The formalization (one of the dimensions of organizational structure) can control or direct employees' behavior. Laws, regulations, job descriptions, and the amount of control by employees can influence employees' perceptions of the richness and meaning of their jobs. The formalization can limit or facilitate relationships between employees. In addition, the formalization can indicate the importance of the staff being familiar with the values and missions of the organization and the degree of attention the organization has to its employees. The centralization determines the extent to which staff and operational managers are empowered to make decisions. The degree of organization centralization, according to the job enrichment theory, is influential in communicating employees with their jobs and the degree of independence and perception of their job richness and meaning. The centralization level of the organization can limit the horizontal or diagonal communication channels of the organization and affect the amount of support staff receive from each other. The centralization level of the organization can also influence employees' judgment of the organization's goals and values and the organization's performance. According to the above, it can be stated that in the long run, the viability and sustainability of an organization's performance will be determined by how real capital will be created from the tangible and intangible assets of the organization in order to satisfy all shareholders. Therefore, the second hypothesis of the study will be as follows:

H2: The leader CQ has a significant effect on the organizational performance enablers

*2.4 Organizational Structural*

To promote and develop soft skills in the organization, a suitable organizational structure is needed. Indeed, Organizational structure is one of the contextual and organizational-influencing factors which provides a foundation for growing intellectual capital [14]. In other words, Ramazan [14] explains that organic organizational structure has a positive impact on intellectual capital, and it enhances intellectual capital in the organization. Balogh, Gaál, and Szabó [37] prove that CQ has a significant effect on the organizational structure. This means that the existence of CQ in the organization leads to changes and improvements in the organizational structure and leads to a structural design that supports the development of talented human resources. The organizational structure is the framework governing the relationships between the jobs, systems and operating processes, and the individuals and groups that strive to achieve the goal. Organizational structure is a set of ways that divides the work into specific tasks and provides coordination between them [38]. The organizational structure must be able to accelerate and facilitate decision making, respond appropriately to the environment, and resolve conflicts between units. The relationship between the core components of an organization and the coordination between its activities and the expression of inter-organizational relationships in terms of reporting are the tasks of the organizational structure [39]. Hence, the third hypothesis of the study will be as follows:

H3: The leader CQ has a significant effect on the organizational structure.

Organizational structure is abundantly mentioned in the literature as a determining factor in organizational performance. For example, Csaszar [40] believes that organizational structure, which shapes decision-making structure, plays a key role in organizational performance. Hunter [41] provides empirical evidence on how elements of organizational structure affect organizational performance. Gaspary, Moura, and Wegner [42] argue that organizational structure is a platform that shapes the organization's performance by facilitating and developing innovation and creativity at work. Therefore, it is interpreted that organizational structure contributes to organizational performance. Accordingly, the fourth hypothesis of this study will be as follows:

H4: The organizational structure has a significant effect on performance enablers.

Many variables are considered as organizational dimensions, but it can be claimed that organizational dimensions are divided into two groups: structural and content [43]. Content dimensions represent the entire organization, such as the size of the organization, the type of technology, its environment, and its objectives. Structural dimensions represent the intrinsic characteristics of an organization and are the basis for measuring and comparing organizations with one another [39]. The organizational structure is the first dimension an organization must design, and then human resources should be hired. Human resources (HR), using the existing form and structure, lead the organization towards a predetermined goal [44]. Therefore, organizational structure influences HR variables. From the structuralist point of view, the societies in which we are born and the institutions, organizations, and groups we belong to are structuring our lives by imposing roles and approaches. According to the structural-functional theory, a structure facilitates and restricts the activities of interacting individuals, and similar activities create structures that facilitate and constrain them [45]. According to Hatch [46], Antony Giddens called the idea "duality of structure" whereby the organizations are facilitated and constrained by structures, procedures, and expectations and, at the same time, form them [46]. Among the structural dimensions of the organization, three factors of complexity, formalization, and centralization, can be identified as the central points of any structure, and the intensity or weakness of each of these three dimensions is effective in the overall formation of the organizational structure. The complexity is measured by the degree of specialization of jobs within the organization as well as by the number of locations in which the organization is located and the number of jobs and organizational positions and hierarchical levels [47]. The formalization indicates the degree of writing, a variety of rules and regulations, and communication practices within the organization [48]. The centralization determines who in the organization has the right to make decisions [47]. The structural model of organizational structures and the criteria to evaluate each dimension are depicted in figure 1.

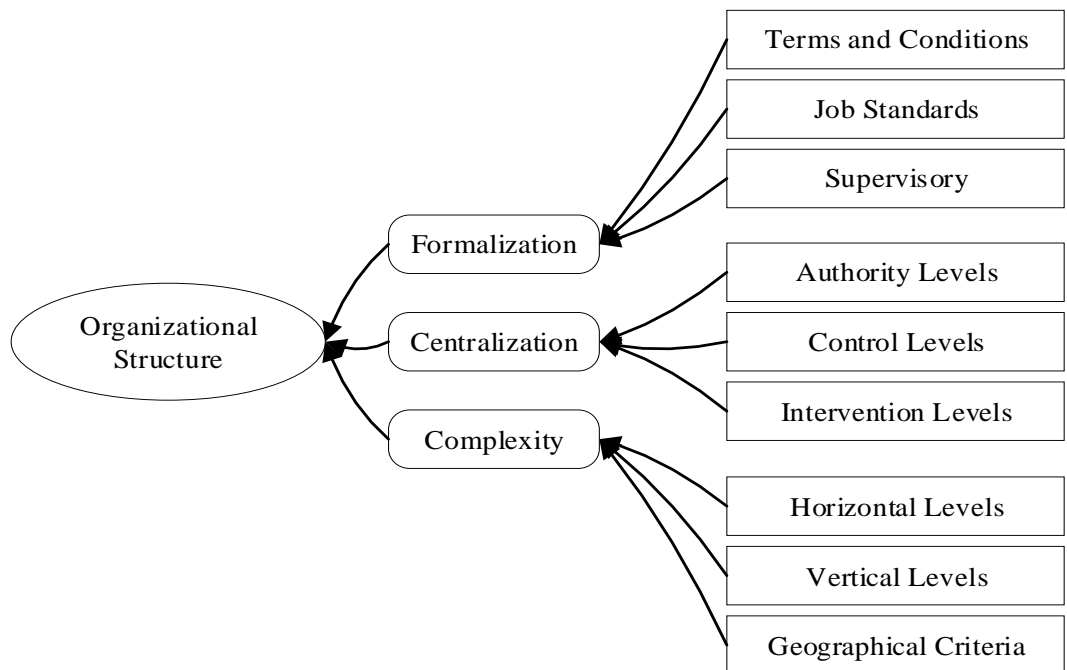

**Figure 1.** Organizational structure dimensions and the evaluation criteria

Source: own compilation based on literature

*2.5 Conceptual model and hypotheses*

In this study, the revised Ang et al. [10] Cultural Intelligence Questionnaire was adopted to investigate leader CQ, after identifying the indicators from the literature. This questionnaire measures CQ in four dimensions: cognitive, meta-cognitive, behavioral, and motivational cultural intelligence. The EFQM standard questionnaire was used for organizational performance variables. This model examines the organization's performance in two areas of enablers and results. Enabling variables are leadership, strategy, people, partnerships and resources, and processes, products, and services. Variables related to the field of results include people's results, customer results, society results, and business results. To measure organizational structure, Robbin's Scale (1987), which measures the three variables of "formalization", "complexity", and "centralization", is used. The proposed conceptual model of the current study is depicted in figure 2.

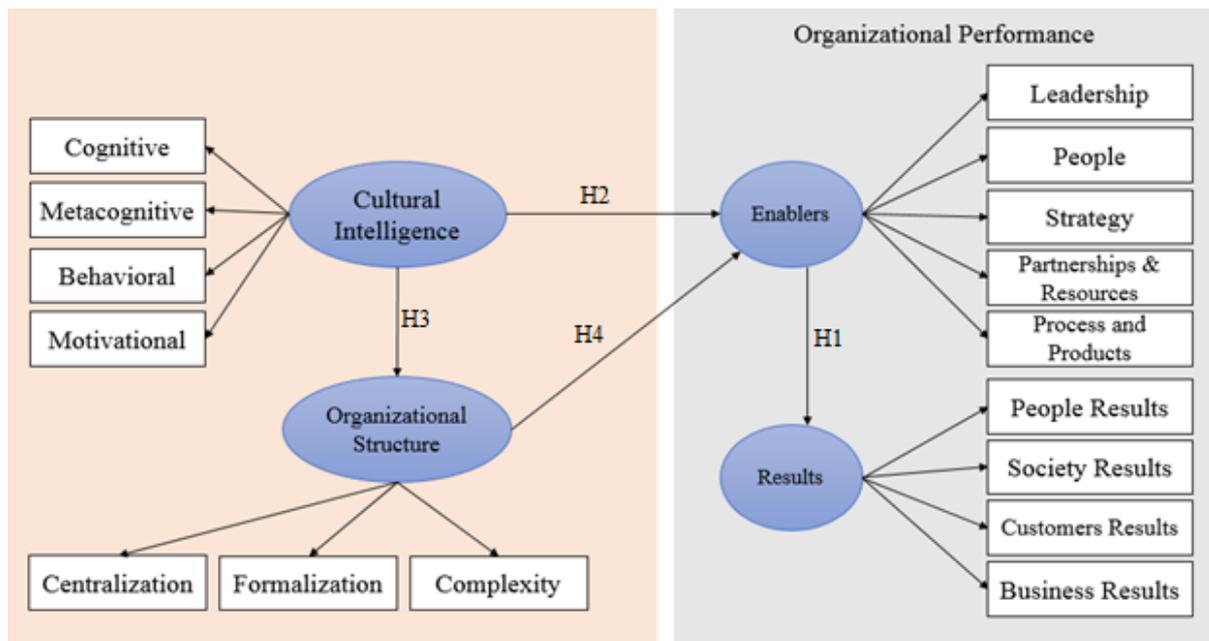

**Figure 2.** Conceptual model of the study

The conceptual model presented in figure 2 is plotted based on the logic of structural equation modeling. As such, leader CQ is considered as a latent variable that is measured in this study using observed variables: cognitive, meta-cognitive, behavioral, and motivational cultural intelligence. The performance enablers are considered as a latent variable that is measured using observed variables: leadership, people, strategy, partnerships and resources, and processes, products, and services. The variable of performance results is also measured in terms of people's results, customer results, business results, and society results. For this reason, the arrays are plotted from the results variable to the observed variables. The organizational structure variables are also measured by formalization, complexity, and centralization. According to the conceptual model, the hypotheses are:

H1: The performance enablers have a significant effect on the performance results.

H2: The leader CQ has a significant effect on the performance enablers.

H3: The leader CQ has a significant effect on the organizational structure.

H4: The organizational structure has a significant effect on performance enablers.

### 3. Methodology

The current research is an applied research in terms of purpose and falls into the category of survey-analytical research. The research method is correlational, where a questionnaire is developed based on the conceptual model of the study, and it is administered for data collection. All managers from a different level of management of knowledge-based companies in Science and Technology Park, Bushehr, Iran, formed the statistical population of the study. It should be noted that due to the limited number of managers, the entire statistical population has been surveyed that was seventy in total. Therefore, seventy questionnaires were distributed among the managers. Of these, fifty-eight questionnaires were identified as suitable for analysis. The validity of the questionnaire was used by experts and academic professors.

The structural equation modeling (SEM) is used to test the hypotheses. Model expression is one of the most important steps in structural equations modeling. In fact, no analysis can be made unless the researcher first expresses the model, which is about the relationships between variables. After model expression, the next step is to obtain the estimation of free parameters from a set of observed data. Iterative methods such as maximum likelihood estimation, generalized least squares, and partial least squares are methods used to estimate a model. The partial least squares method, also introduced in the discussion of regression modeling with PLS, is one of the multivariate statistical methods that can be used, despite some limitations (e.g. uncertainty of the response variable distribution, the low observations, or the existence of a serious correlation between the explanatory variables), to model one or more response variables simultaneously for several explanatory variables. Due to the low sample size and the non-normality of the response variable distribution, the PLS using SmartPLS 3.2.8 was used to analyze the data and test the hypotheses.

The SEM includes two parts: 1) structural model and 2) measurement models. A structural model refers to the relationship between the latent variables or the concepts. It is worth mentioning that the latent variables are those variables that cannot be measured directly. Therefore, the measurement models are applied. Indeed, a latent variable is measured by observed variables. Thus, the measurement models comprise relationships between the observed variables and a corresponding latent variable. There are mainly two types of measurement models: the reflective model, the formative model. Since the measurement model of the current study is a reflective model, this type of measurement model is explained in this study. In modeling, the direction of the arrows is outward the latent variable. In a reflective model, the observed variables reflect the corresponding latent variable in the form of the regression model as follow:

$$x_{pq} = \lambda_{p0} + \lambda_{pq}\xi_q + \epsilon_{pq} \qquad (1)$$

Where $\lambda$ refers to the loading factor, and $\xi$ is the latent variable, and $\epsilon$ is the error in the measurement process.

A reflective model is indeed a factor analysis model. Since the current study is an explorative study, the accuracy and adequacy of the sample play an important rule in testing the model fit. Therefore, the Kaiser-Meyer-Olkin (KMO) is applied to test the adequacy of the sample (not the sample size). KMO is typically used to checking the adequacy of the sample in the explorative factor analysis. The value of KMO is between 0 to 1, where values close to 1 disclose the sum of the correlations is higher than the partial correlations, which represent a good fit, and the values higher than 0.6 consider suitable. Equation 1 shows how KMO is measured.

$$KMO = \frac{\sum_{i \neq j} r_{ij}^2}{\sum_{i \neq j} r_{ij}^2 + \sum_{i \neq j} u_{ij}^2} \qquad (2)$$

Where $r_{ij}$ stands for the correlation matrix, and $u_{ij}$ represents the partial covariance matrix.

The average variance extracted (AVE) indicates the degree of correlation of a reflective model, and it tests the convergent validity. In other words, this measurement shows the amount of variance that a latent variable capture from its observed variables in comparison with the amount of variance gets from the error measurement. The greater the correlation, the higher the fit. The acceptable criterion for convergent validity is the figures higher than 0.7 are very good, and the figures higher than 0.5 are acceptable which represents more than 50% variance of the structure should be covered by its own indicators, and it can be calculated as follows:

$$AVE = \frac{\sum \lambda_i^2}{\sum \lambda_i^2 + \sum \Theta_{ij}} \qquad (3)$$

Where $\lambda_i^2$ is the factor loading and $\Theta_{ij}$ is error variance and can be calculated as follow:

$$\Theta ij = \sum 1 - \lambda_i^2 \qquad (4)$$

Internal consistency is significant issue in the reflective models too. The composite reliability ($\rho c$) measures internal consistency, and it uses along with Cronbach's alpha to test the reliability, and it is evaluated as follows:

$$\rho c = \frac{(\sum \lambda_i)^2}{(\sum \lambda_i)^2 + \sum \Theta_{ij}} \qquad (5)$$

As it is mentioned above, the structural model refers to the relationship among the latent variables. In the SEM, the dependent variable is called an endogenous latent variable, and the independent variable is named exogenous latent variable. Follow the equation for the calculation of an endogenous latent variable is provided.

$$\xi = \beta_{oj} + \sum \beta_{qj} \xi_q + \zeta_j \qquad (6)$$

Where $\xi$ is an endogenous latent variable, $\beta_{qj}$ is the path coefficient between the $q$ exogenous latent variable. j represents the endogenous variable. $\zeta_j$ refers to the error in the inner relation.

### 4. Results

In this section, firstly, the results related to the demographic features of the studied sample are described and then the results of model testing and hypotheses are provided. As it is presented in Table 1, 58.6% of the respondents were female, and 41.4% were male. In terms of education, 4.4% were diplomas, 12.2% associate degrees, 68.3% bachelors, 14.4% masters and 0.6% PhDs. The data related to the work experience of the participants in this study show that the work experience of 13 participants is under five years, the work experience of 17 of them is between 5 and 10 years, and the work experience of 28 participants is over ten years.

**Table 1.** Demographic features of the study sample

| Demographic Characteristics | | Frequency | Percentage |
|---|---|---|---|
| **Gender** | Male | 24 | 41.4% |
| | Female | 34 | 58.6% |
| **Total** | | 58 | 100% |
| **Experience** | 0-5 | 13 | 23% |
| | 5-10 | 17 | 29% |
| | Over 10 | 28 | 48% |
| **Total** | | 58 | 100% |
| **Academic Qualifications** | Diploma | 2 | 3% |
| | Associated Degree | 7 | 12% |
| | Bachelor's Degree | 40 | 69% |
| | Master's Degree | 8 | 14% |
| | Ph.D. | 1 | 2% |
| **Total** | | 58 | 100% |

*4.1 Conceptual Model Testing*

The partial least squares structural equation modeling is employed for model testing. Before all, to test the sample accuracy, the KMO metrics is done. According to Table 2, the KMO related to each of the variables is higher than 0.6, that refers to the acceptable accuracy of the sample of the study. It is worth mentioning that the validity of the questionnaire is confirmed by the university professors who are experts in the related fields. Besides, for evaluation of convergent validity, the measurement of AVE is tested, and the results are summarized in Table 2 as well. Table 2 indicates that the AVE value for all the variables is higher than 0.5 which implies more than 50% variance of each structure (variables) are covered by its own indicators, which is desirable. Cronbach's alpha and the composite reliability (CR) also are measured to test the reliability of the model measurement model (the questionnaire). According to the results, The Cronbach's alpha for all three variables of cultural intelligence (0.92), organizational performance (0.87), and organizational structure (0.84) is higher than 0.7 and in the acceptable range. The value of CR also for all the variables is higher than 0.7, which indicates the acceptable reliability of the questionnaire (see Table 2).

**Table 2:** Validity and Reliability of the variables

| Variables | KMO | AVE | Cronbach's alpha | CR |
|---|---|---|---|---|
| Cultural Intelligence | 0.62 | 0.722 | 0.92 | 0.933 |
| Organizational Performance | 0.91 | 0.789 | 0.87 | 0.922 |
| Organizational Structure | 0.83 | 0.672 | 0.84 | 0.918 |

Figure 3 is the output of testing the model in SmartPLS 3.2.8. the numbers inside the ellipse are the coefficient of determination ($R^2$). The $R^2$ determines how many percents of the variations of a dependent variable is explained by the independent variable/s [49]. Therefore, it is evident that this value is equal to zero for the independent variable and higher than zero for the dependent variable. The higher the $R^2$ value, the more considerable influence of the independent variable on the dependent variable. According to the coefficient of determination of the model, it is interpreted that all dimensions of cultural intelligence (cognitive, meta-cognitive, behavioral, and motivational cultural intelligence) were able to explain 72.2% of the variance of performance enablers and 49.1% of the difference of organizational structure, and performance enablers are able to explain 85.5% of the variance of performance results. Residuals are related to prediction error and may include other factors influencing the dimensions of performance variables (enablers and effects) and organizational structure.

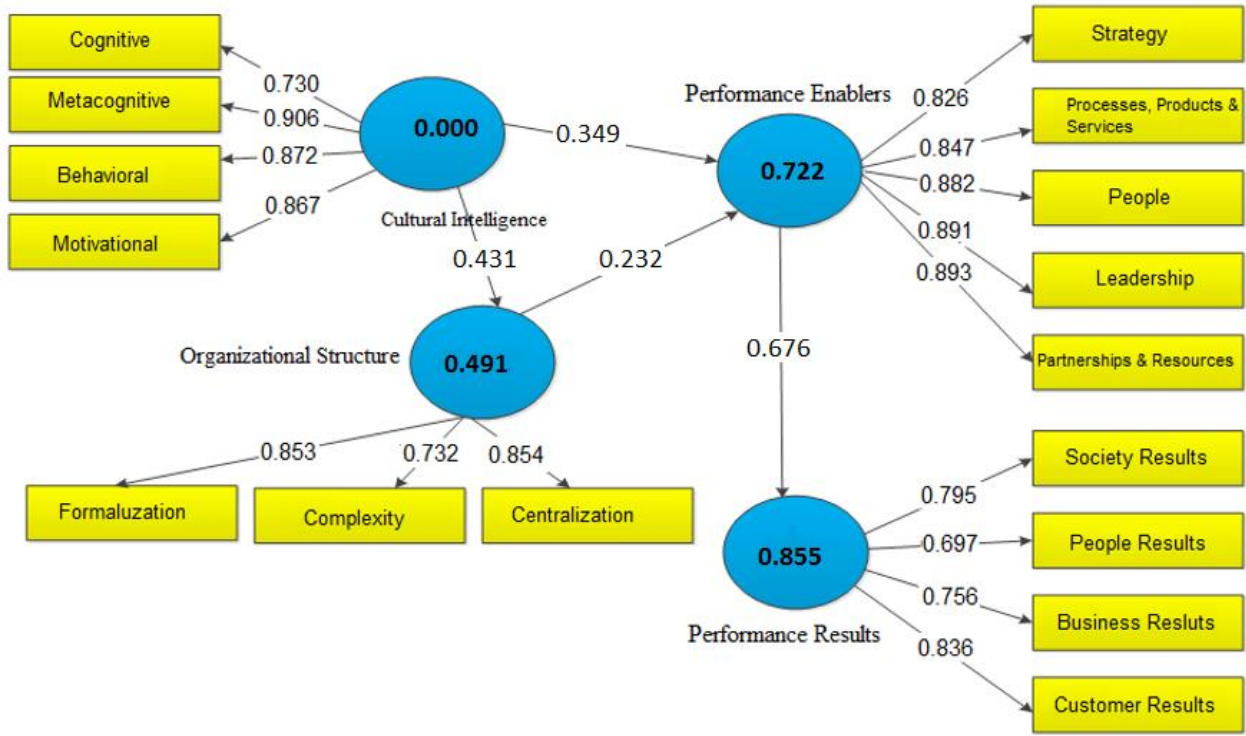

**Figure 3.** The test of the proposed model of the study; R-squares and Path Coefficients

Figure 4 illustrates the model of the effect of the cultural intelligence of leaders on organizational performance through the mediating variable of organization structure in absolute value coefficient (|t-value|). This model, in fact, tests all measurement equations (loadings factor) and structural equations (path coefficients) using t-statistic. According to this model, the path coefficients and loading factors are significant at a 95% confidence level; if the t-value is either higher than 1.96 and or less than 1.96, then the corresponding loading factor or path coefficient is not significant. In addition, the path coefficients and loading factors are significant at a 99% confidence level, providing that the corresponding t-value is higher than 2.58.

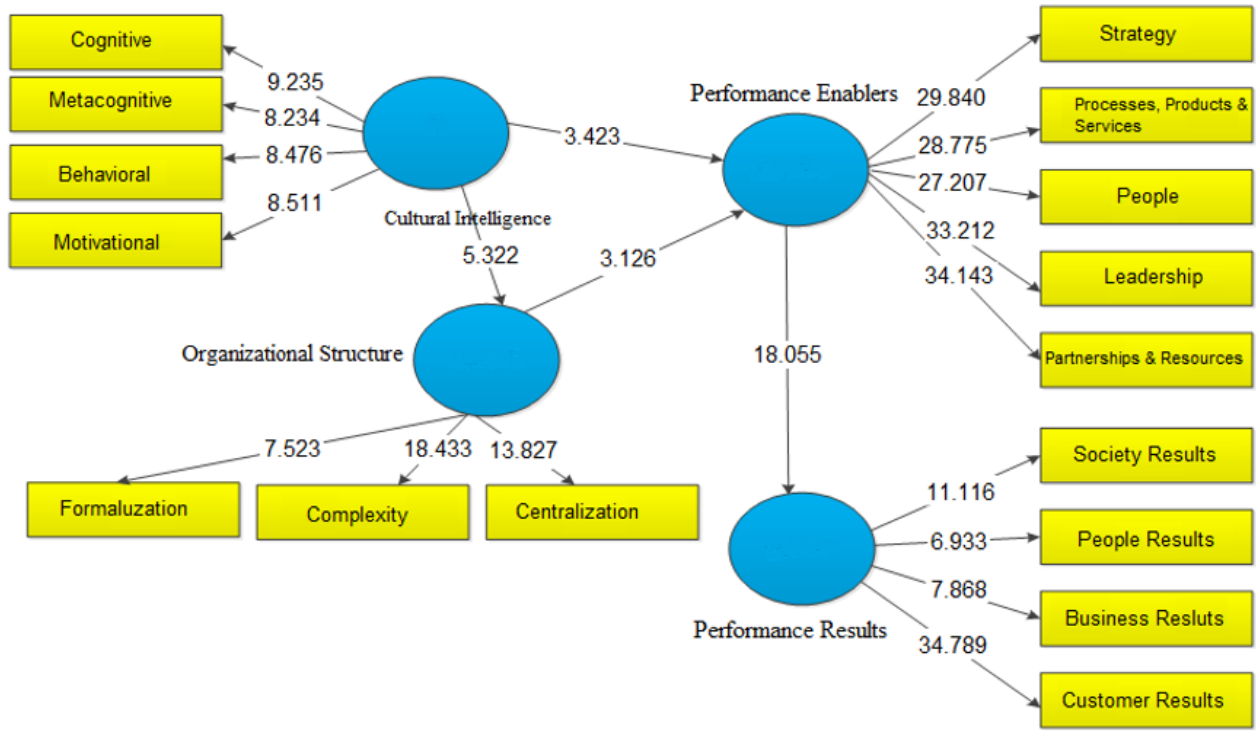

**Figure 4.** The conceptual model of the study and the results of |t-value| test

Confirmatory factor analysis was used to structure the questionnaire and identify the constituent factors of each construct. The results of the confirmatory factor analysis of structures are summarized in Table 3. All the loading factors were tested at two levels of 5% and 1% significant level. According to the result, all the loading factors were significant at the 99% confidence level and were able to make a substantial contribution to the measurement of the relevant structures.

**Table 3.** The loading factors of observed variables

| Observable Variables | Leader CQ | Enablers | Results | Organizational Structure |
|---|---|---|---|---|
| Cognitive | 0.730 | | | |
| Metacognitive - | 0.906 | | | |
| Motivational | 0.872 | | | |
| Behavioral | 0.867 | | | |
| Leadership | | 0.891 | | |
| Strategy | | 0.826 | | |
| Partnerships and resources | | 0.893 | | |
| Processes, Products, and services | | 0.847 | | |
| People | | 0.882 | | |
| Customer Results | | | 0.836 | |
| People Results | | | 0.679 | |
| Business Results | | | 0.756 | |
| Society Results | | | 0.795 | |
| Formalization | | | | 0.843 |
| Complexity | | | | 0.730 |
| Centralization | | | | 0.532 |

** all the loading factors are significant at the 99% confidence level

After examining the fit of measurement models, structural models, and a general model, the hypotheses are tested. According to Table 4, all the hypotheses of the study confirmed. Since the path coefficient of the cultural intelligence to the performance enablers ($\beta=0.349$) is significant at a 99% confidence level. Therefore, the respective $R^2$ is acceptable, that implies the first hypothesis is confirmed. It means that leader CQ positively affects the performance enablers. In addition, the path coefficient of the leaders' cultural intelligence to the organizational structure (β=0.431) is also significant at a 99% confidence level that indicates the second hypothesis of the study confirmed as well. In other words, there is no evidence to refuse the positive impact of leader CQ on the organizational structure. Likewise, the importance of the role of organizational structure in the performance enablers is confirmed as well. As it is provided in Table 4, the path coefficient of the organizational structure and the performance enablers is equal to 0.232, and it is significant at a 99% confidence level. Ultimately, the path coefficient of the organizational performance enablers to the performance results (β=0.676) is also significant at 99% confidence level that implies the performance enablers affect the performance results positively and according to the $R^2$ results, 85.5% of the performance results are explained by the performance enablers that is a considerable number.

Table 4. The results of hypotheses testing

| Hypotheses | Path Coefficient (β) | T-value | $R^2$ | Result |
|---|---|---|---|---|
| Cultural Intelligence → Performance enablers | 0.349** | 3.423 | 0.722 | Confirmed |
| Cultural Intelligence → Organizational Structure | 0.431** | 5.322 | 0.491 | Confirmed |
| Organizational Structure → Performance enablers | 0.232** | 3.126 | 0.722 | Confirmed |
| Performance enablers → Performance Results | 0.676** | 18.055 | 0.855 | Confirmed |

** the correspond path coefficient is significant at 99% confidence level

## 5. Discussion

This study investigated the determinant role of leader CQ in organizational performance. In this study, the mediating role of organizational structure in explanation of organizational performance is also examined. Since there is no study in this area the results obtained from the current study cannot be compared with other studies, only the findings of this study can be compared with the concepts provided by the experts in the fields of cultural intelligence and organizational performance which were discussed and compared in the research background section of the current study. The results of this study revealed that leader CQ directly and indirectly (i.e., through organizational structure) affects organizational performance. Although there is no study in the literature to examine the effect of leader CQ on organizational performance, there is evidence in the literature proving the importance of leader CQ in leaders' performance and team performance (e.g., [10-13]) and the results of the current study is in accordance with these studies. The results of testing the first hypothesis of the present study, which examines the relationship between performance enablers and performance results, showed that enablers have a positive and significant effect on organizational performance results, which is consistent with the EFQM model. The second hypothesis emphasized that the variable of leader CQ had a significant effect on performance enablers. This hypothesis was accepted based on the results and it showed that the mentioned variable had a positive and significant effect on the enablers and this relationship was accepted with 95% confidence. This indicates that any attempts in direction of impowering the CQ components, which are the cognitive intelligence, behavioral intelligence, motivational intelligence, and meta-cognitive intelligence, empower the performance enablers. The third hypothesis declared that cultural intelligence of leaders has a significant effect on organizational structure. This hypothesis was accepted based on the findings and the results showed that the mentioned variable has a positive and significant effect on the organizational structure and this relationship was accepted with 95% confidence. It means that the leader CQ level determines the organizational structure. However, in the present study, the types of organizational structures and their relationship with different levels of cultural intelligence of leaders have not been studied. The fourth hypothesis tested the importance of the role of organizational structure in the performance enablers. This hypothesis is also confirmed at 99% confidence level. In other words, organizational structure affects organizational performance and different organizational structures lead to different performance results. This finding is also consistent with the findings of Csaszar [40], Hunter [41] and Gaspary et al. [42]. CQ is recognized as an ability to communicate effectively with people from different cultures and subcultures. In other words, today's leaders and staff must have the flexibility to consciously adapt to any new cultural situation they face. In this regard, it is the CQ and empowerment of individuals and managers that will rally to the aid of the organization. Because culturally unintelligent people may not be able to communicate with their colleagues from the same or other cultures, they may have difficulty in understanding their business. Managers and supervisors who ignore the impact of international cultures on decision-making will fail in an attempt to improve quality unless they align development with development of culture. In contrast, culturally intelligent individuals are able to interpret the behavior of others and, even if necessary, adapt to the behavior of others.

## 6. Conclusion

The main objective of the current study was to investigate the impact of leader CQ on organizational performance mediated by organizational structure among knowledge-based companies in the Science and Technology Park, Bushehr, Iran. The results of this study bridge theoretical gaps in the literature. First, current literature is limited to studies that examine the leader CQ impact on either the leader's performance or the team's performance, and there is no research to examine the impact of the leader CQ on organizational performance. The present study showed that leader CQ affects the performance enablers and the higher CQ of leaders, the higher the performance of that organization. Second, this study considers organizational structure as the mediate variable facilitating the impact of leader CQ on the organizational performance. Organizational structure was used as an intermediary variable because organizations need a suitable structure to nurture and use their intellectual capital. The findings showed that leader CQ affects organizational structure and organizational structure in turn affects organizational performance. Such an approach enables this research to theorize the interrelationship among these three variables, naming leader CQ, organizational structure, and organizational performance, and proposes a model to enhance the organizational performance. Third, despite the proposed model of the study is examined in Iran, where the context is different than other developed western countries, the results are consistent with other results in the literature. This implies that leader CQ, which is a soft skill that allows leaders to communicate effectively in a multicultural environment, is not related to cultures and is a skill required for the leader of companies operating in multicultural environments, regardless of geographical boundaries. Once the diversity of all aspects of human life is fully embraced, the need for effective intercultural leaders is increasingly felt. One of the most important attributes of such leaders is undoubtedly the ability to manage increasing cultural diversity. In the meantime, it is important to pay attention to CQ and to strive to improve it. CQ is the ability to learn new patterns in cultural interactions and provide correct behavioral responses to these patterns. With the expansion of international business activities, empowering managers is needed to cope with leading cultural complexities. Considering the results and the effect of leader CQ on organizational performance, it can be stated that investing on the human capital specially leader CQ in the workplace is one of the fundamental factors that increase the performance of organizations.

There are limitations for generalization of the findings of the current study. This study was taken place across the knowledge-based companies in the Science and Technology Park in a city in Iran. Therefore, the findings do not represent all knowledge-based companies in Iran, nor do they represent other forms of companies in Iran. This study showed that leader CQ affects organizational structure and through this effect, leader CQ also indirectly affects organizational performance. However, in this study, organizational structures tailored to leaders with different levels of CQ have not been studied. Therefore, for future studies, it is recommended to determine what kind of organizational structure the leaders with different levels of CQ prefer. In addition, it is also recommended that for future research to test the proposed model of the current study among organizations in other industries and even other countries and compare the results with the results of the present study.

References

1. Morley, Michael J, Jean-Luc Cerdin, and Taewon Moon. "Emotional Intelligence Correlates of the Four-Factor Model of Cultural Intelligence." *Journal of Managerial Psychology* (2010).
2. Thomas, David C. "Cultural Intelligence." *Wiley Encyclopedia of Management* (2015): 1-3.
3. Elrehail, Hamzah, Okechukwu Lawrence Emeagwali, Abdallah Alsaad, and Amro Alzghoul. "The Impact of Transformational and Authentic Leadership on Innovation in Higher Education: The Contingent Role of Knowledge Sharing." *Telematics and Informatics* 35, no. 1 (2018): 55-67.
4. Alzghoul, Amro, Hamzah Elrehail, Okechukwu Lawrence Emeagwali, and Mohammad K AlShboul. "Knowledge Management, Workplace Climate, Creativity and Performance." *Journal of Workplace Learning* (2018).
5. Velten, Laura, and Conrad Lashley. "The Meaning of Cultural Diversity among Staff as It Pertains to Employee Motivation." *Research in Hospitality Management* 7, no. 2 (2018): 105-13.
6. Wu, Wei-Wen, and Yu-Ting Lee. "Developing Global Managers' Competencies Using the Fuzzy Dematel Method." *Expert systems with applications* 32, no. 2 (2007): 499-507.
7. Harvey, Michael, Milorad M Novicevic, and Timothy Kiessling. "Development of Multiple Iq Maps for Use in the Selection of Inpatriate Managers: A Practical Theory." *International Journal of Intercultural Relations* 26, no. 5 (2002): 493-524.
8. Bahrami, Parvaneh, Saeed Nosratabadi, and Csaba Bálint Illés. "Role of Intellectual Capital in Corporate Entrepreneurship." *Calitatea* 17, no. 155 (2016): 111.
9. Alon, Ilan, and James M Higgins. "Global Leadership Success through Emotional and Cultural Intelligences." *Business horizons* 48, no. 6 (2005): 501-12.
10. Ang, Soon, Linn Van Dyne, Christine Koh, K Yee Ng, Klaus J Templer, Cheryl Tay, and N Anand Chandrasekar. "Cultural Intelligence: Its Measurement and Effects on Cultural Judgment and Decision Making, Cultural Adaptation and Task Performance." *Management and organization review* 3, no. 3 (2007): 335-71.
11. Rosenauer, Doris, Astrid C Homan, Christiane AL Horstmeier, and Sven C Voelpel. "Managing Nationality Diversity: The Interactive Effect of Leaders' Cultural Intelligence and Task Interdependence." *British Journal of Management* 27, no. 3 (2016): 628-45.
12. Dogra, AS, and Varsha Dixit. "Cultural Intelligence: Exploring the Relationship between Leader Cultural Intelligence, Team Diversity and Team Performance." *ELK Asia Pacific Journal of Human Resource Management and Organisational Behaviour* 3, no. 1 (2016).
13. Groves, Kevin S, and Ann E Feyerherm. "Leader Cultural Intelligence in Context: Testing the Moderating Effects of Team Cultural Diversity on Leader and Team Performance." *Group & Organization Management* 36, no. 5 (2011): 535-66.

14. Ramezan, Majid. "Intellectual Capital and Organizational Organic Structure in Knowledge Society: How Are These Concepts Related?" *International Journal of Information Management* 31, no. 1 (2011): 88-95.
15. Bahrami, Parvaneh, Saeed Nosratabadi, and B Cs Illés. "Effects of Intellectual Capital Components on Corporate Entrepreneurship." Paper presented at the Proceedings of the 1th International Conference Contemporary Issues in the Theory and Practice of Management 2016.
16. Kaufmann, Wesley, Erin L Borry, and Leisha DeHart-Davis. "More Than Pathological Formalization: Understanding Organizational Structure and Red Tape." *Public Administration Review* 79, no. 2 (2019): 236-45.
17. Gentile-Lüdecke, Simona, Rui Torres de Oliveira, and Justin Paul. "Does Organizational Structure Facilitate Inbound and Outbound Open Innovation in Smes?" *Small Business Economics* (2019): 1-22.
18. Sandhu, Sukhbir, and Carol T Kulik. "Shaping and Being Shaped: How Organizational Structure and Managerial Discretion Co-Evolve in New Managerial Roles." *Administrative Science Quarterly* (2018): 0001839218778018.
19. Earley, P Christopher, and Soon Ang. *Cultural Intelligence: Individual Interactions across Cultures*: Stanford University Press, 2003.
20. Ang, Soon, Linn Van Dyne, and Christine Koh. "Personality Correlates of the Four-Factor Model of Cultural Intelligence." *Group & Organization Management* 31, no. 1 (2006): 100-23.
21. Charoensukmongkol, Peerayuth. "Cultural Intelligence and Export Performance of Small and Medium Enterprises in Thailand: Mediating Roles of Organizational Capabilities." *International Small Business Journal* 34, no. 1 (2016): 105-22.
22. Ramsey, Jase R, Jordan Nassif Leonel, Geovana Zoccal Gomes, and Plinio Rafael Reis Monteiro. "Cultural Intelligence's Influence on International Business Travelers' Stress." *Cross Cultural Management: An International Journal* 18, no. 1 (2011): 21-37.
23. Ramirez, Andrea Reyes. "Impact of Cultural Intelligence Level on Conflict Resolution Ability: A Conceptual Model and Research Proposal." *Emerging Leadership Journeys* 3, no. 1 (2010): 42-56.
24. Gregory, Robert, Michael Prifling, and Roman Beck. "The Role of Cultural Intelligence for the Emergence of Negotiated Culture in It Offshore Outsourcing Projects." *Information Technology & People* 22, no. 3 (2009): 223-41.
25. Earley, P Christopher, and Heidi K Gardner. "Internal Dynamics and Cultural Intelligence in Multinational Teams." In *Managing Multinational Teams: Global Perspectives*, 3-31: Emerald Group Publishing Limited, 2005.
26. Shin, Duckjung, and Alison M Konrad. "Causality between High-Performance Work Systems and Organizational Performance." *Journal of Management* 43, no. 4 (2017): 973-97.
27. Asree, Susita, Mohamed Zain, and Mohd Rizal Razalli. "Influence of Leadership Competency and Organizational Culture on Responsiveness and Performance of Firms." *International Journal of Contemporary Hospitality Management* 22, no. 4 (2010): 500-16.
28. Oggioni, G, R Riccardi, and R Toninelli. "Eco-Efficiency of the World Cement Industry: A Data Envelopment Analysis." *Energy Policy* 39, no. 5 (2011): 2842-54.
29. Young, Chaur-Shiuh, Hwan-Yann Su, Shih-Chieh Fang, and Shyh-Rong Fang. "Cross-Country Comparison of Intellectual Capital Performance of Commercial Banks in Asian Economies." *The Service Industries Journal* 29, no. 11 (2009): 1565-79.

30. Sozuer, Aytug. "Self Assessment as a Gate to Performance Improvement: A Study on Hospitality Management in Turkey." *Procedia-Social and behavioral sciences* 24 (2011): 1090-97.
31. Gopalakrishna, Pradeep, and Mahesh Chandra. "Malcolm Baldrige, Deming Prize and European Quality Awards: A Review and Synthesis." In *Handbook of Total Quality Management*, 755-69: Springer, 1998.
32. Berssaneti, Fernando Tobal, Ana Maria Saut, Májida Farid Barakat, and Felipe Araujo Calarge. "Is There Any Link between Accreditation Programs and the Models of Organizational Excellence?" *Revista da Escola de Enfermagem da USP* 50, no. 4 (2016): 650-57.
33. Sampaio, Paulo, Pedro Saraiva, and Ana Monteiro. "A Comparison and Usage Overview of Business Excellence Models." *The TQM Journal* 24, no. 2 (2012): 181-200.
34. Escrig-Tena, Ana B, Beatriz Garcia-Juan, and Mercedes Segarra-Ciprés. "Drivers and Internalisation of the Efqm Excellence Model." *International Journal of Quality & Reliability Management* 36, no. 3 (2019): 398-419.
35. Plum, Elizabeth. "Cultural Intelligence: A Concept for Bridging and Benefiting from Cultural Differences." *"Glocal" working* (2007): 80.
36. Bibikova, A, and V Kotelnikov. "Cultural Intelligence (Cq): Knowledge, Arts and Skills." 2006.
37. Balogh, Agnes, Zoltán Gaál, and Lajos Szabó. "Relationship between Organizational Culture and Cultural Intelligence." *Management & Marketing* 6, no. 1 (2011): 95.
38. Ahmady, Gholam Ali, Maryam Mehrpour, and Aghdas Nikooravesh. "Organizational Structure." *Procedia-Social and behavioral sciences* 230 (2016): 455-62.
39. Ashkenas, Ron, Dave Ulrich, Todd Jick, and Steve Kerr. *The Boundaryless Organization.: Breaking the Chains of Organizational Structure*: John Wiley & Sons, 2015.
40. Csaszar, Felipe A. "Organizational Structure as a Determinant of Performance: Evidence from Mutual Funds." *Strategic Management Journal* 33, no. 6 (2012): 611-32.
41. Hunter, Judy. "Improving Organizational Performance through the Use of Effective Elements of Organizational Structure." *Leadership in Health Services* (2002).
42. Gaspary, Eliana, Gilnei Luiz De Moura, and Douglas Wegner. "How Does the Organisational Structure Influence a Work Environment for Innovation?" *International Journal of Entrepreneurship and Innovation Management* 24, no. 2-3 (2020): 132-53.
43. Young, Courtney, and Sumantra Ghoshal. *Organization Theory and the Multinational Corporation*: Springer, 2016.
44. Lepak, Dave, Kaifeng Jiang, and Robert E Ployhart. "Hr Strategy, Structure and Architecture." In *A Research Agenda for Human Resource Management*: Edward Elgar Publishing, 2017.
45. Kitchen, Deeb Paul. "Structural Functional Theory." *Encyclopedia of Family Studies* (2016): 1-7.
46. Hatch, Mary Jo. *Organization Theory: Modern, Symbolic, and Postmodern Perspectives*: Oxford university press, 2018.
47. Fry, Louis W, and John W Slocum Jr. "Technology, Structure, and Workgroup Effectiveness: A Test of a Contingency Model." *Academy of management journal* 27, no. 2 (1984): 221-46.
48. Marsh, Robert M, and Hiroshi Mannari. "Technology and Size as Determinants of the Organizational Structure of Japanese Factories." *Administrative Science Quarterly* (1981): 33-57.

49. Ringle, Christian, Dirceu Da Silva, and Diógenes Bido. "Structural Equation Modeling with the Smartpls." *Bido, D., da Silva, D., & Ringle, C.(2014). Structural Equation Modeling with the Smartpls. Brazilian Journal Of Marketing* 13, no. 2 (2015).